\documentstyle[epsfig,12pt]{article}
\newcommand{\case}[2]{\mbox{\footnotesize $\displaystyle \frac{#1}{#2}$}}

\begin{document}
{\Large\bf Differences between heavy and light quarks}\\[2mm]
{\it P. Maris and C. D. Roberts}\\[2mm]
{\small Physics Division, 
Argonne National Laboratory, Argonne IL 60439-4843}\\[2mm]
{\bf Abstract} 

The quark Dyson--Schwinger equation shows that there are distinct
differences between light and heavy quarks. The dynamical mass function
of the light quarks is characterised by a sharp increase below $1$~GeV,
whereas the mass function of the heavy quarks is approximately constant
in this infrared region. As a consequence, the heavy-meson masses
increase linearly with the current-quark masses, whereas the light
pseudoscalar meson masses are proportional to the square root of the
current-quark masses.
\smallskip

{\bf Quark propagator}

The Dyson--Schwinger equations [DSEs] provide an excellent tool to study
nonperturbative aspects of quark propagators and bound states$^1$.
Various properties of the light mesons can be described very well in
this framework$^2$ but this method is not limited to the light quark
sector. Nonperturbative effects, such as chiral symmetry breaking and
confinement, are reflected in the dressed quark propagator $S(p)^{-1} =
i \gamma\cdot p A(p^2) + B(p^2)$. In Euclidean space, the DSE for the
quark propagator is
\begin{eqnarray}
\label{gendse}
S(p)^{-1} & = & Z_2 i\gamma\cdot p + Z_4 m(\mu) 
        + \, Z_1\, \int\frac{{\rm d}^4q}{(2\pi)^4}\,
        g^2 D_{\mu\nu}(p-q) {\hbox{$\frac{\lambda^a}{2}$}} \gamma_\mu S(q)
        \Gamma^a_\nu(q,p) \,,
\end{eqnarray}
where $D_{\mu\nu}(k)$ is the dressed gluon propagator, $\Gamma^a_\nu(q;p)$
is the dressed quark-gluon vertex and $m(\mu)$ is the renormalised
current-quark mass. The quark-gluon-vertex, quark wave-function, 
and quark mass renormalisation constants, $Z_1$, $Z_2$, and $Z_4$
respectively, depend on the renormalisation point $\mu$, as does 
the current-quark mass.

The set of DSEs form an infinite hierarchy of integral equations, and
for actual calculations we need to truncate this set of equations.
Following Ref.~1, we use a truncation based on the so-called ladder 
approximation, namely
\begin{equation}
\label{ouransatz}
 Z_1\, \Gamma^a_\nu(q,p)
 g^2 D_{\mu\nu}(p-q) {\hbox{$\frac{\lambda^a}{2}$}} \gamma_\mu 
\to
 {\hbox{$\frac{\lambda^a}{2}$}}\gamma_\nu 
 {\cal G}((p-q)^2)\, D_{\mu\nu}^{\rm free}(p-q)
 {\hbox{$\frac{\lambda^a}{2}$}}\gamma_\mu \,.
\end{equation}
Qualitatively reliable studies of the DSE for the dressed gluon
propagator indicate that the effective interaction is significantly
enhanced in the infrared such that on this domain it is well represented
by an integrable singularity$^3$. Combining this observation with
asymptotic freedom in the ultraviolet, motivates an Ansatz$^2$ for
${\cal G}((p-q)^2)$ with a $\delta-$function to represent the integrable
infrared singularity, plus a tail term, regular in the infrared region,
to incorporate asymptotic freedom. The renormalisation constants are
fixed by applying the renormalisation condition $A(\mu^2) = 1$ and
$B(\mu^2) = m(\mu)$.

Now that we have fully specified the truncation of the quark DSE, we can
solve numerically the resulting coupled integral equations for 
$A(p^2)$ and $B(p^2)$. In Fig.~1, we show the dynamical quark
mass function $M(p^2)=B(p^2)/A(p^2)$ obtained for different
quark flavours. 
\begin{figure}[t]
\vspace*{-5mm}

\hspace*{-0.7cm}
\vspace*{-7.5cm}\epsfig{figure=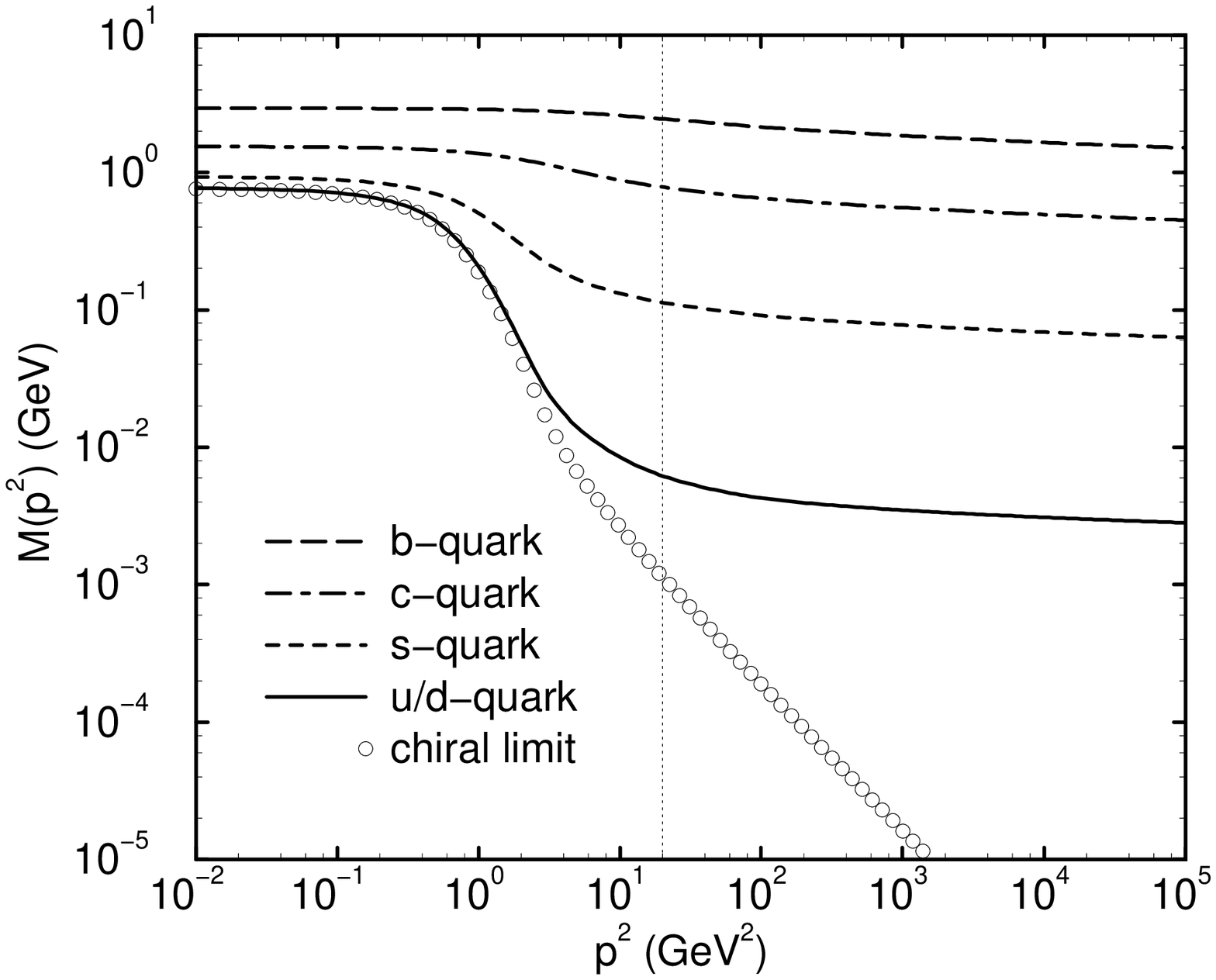,height=7.5cm}

\hspace*{7.6cm}\epsfig{figure=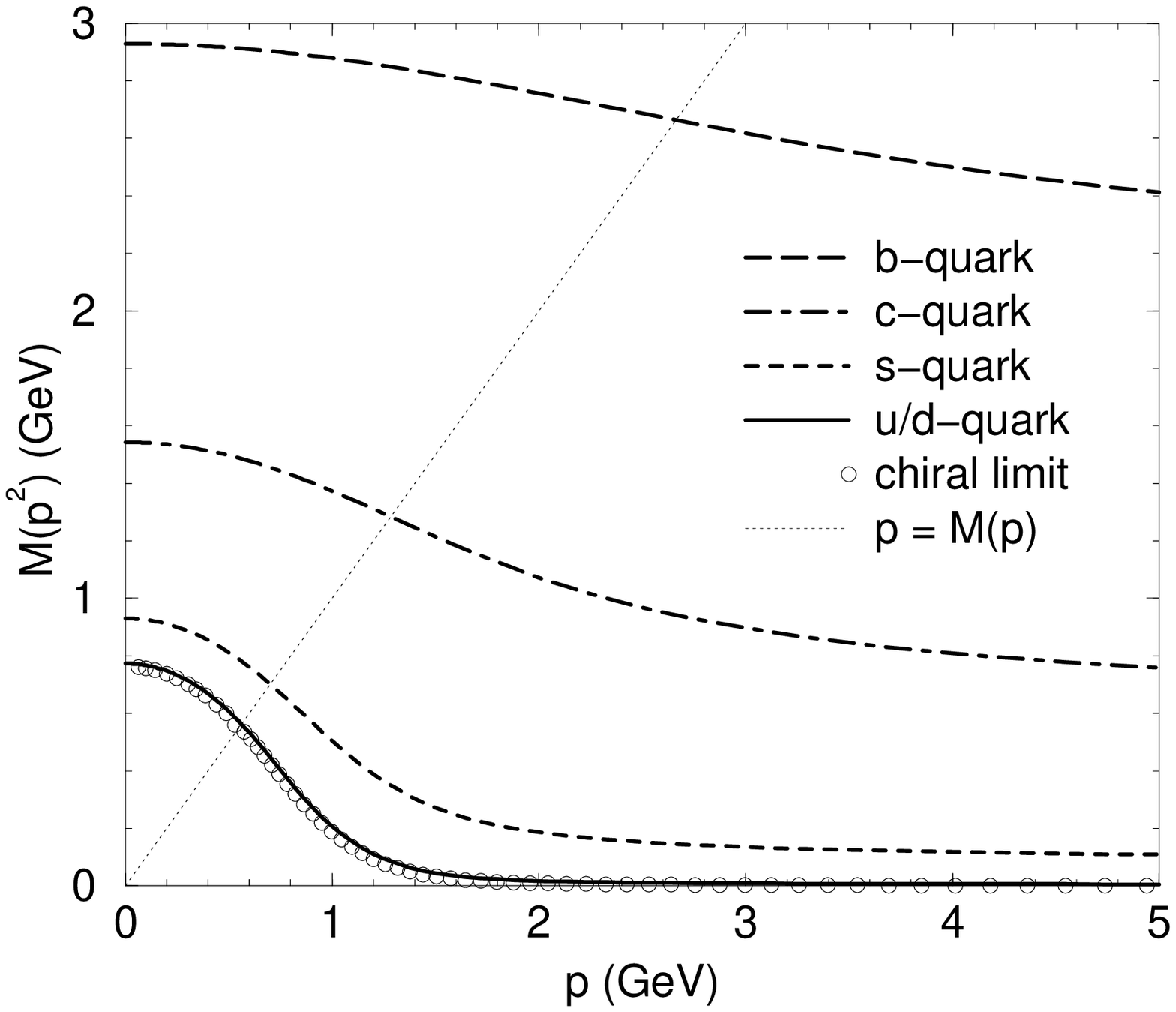,height=7.5cm}

\vspace*{-1cm}

{\small Fig.~1. The dynamical mass function of the quark propagator in
the chiral limit (circles), and for up/down (solid), strange (short
dash), charm (dash-dot), and bottom (long dash) quarks.}
\end{figure}
With a nonzero current-quark mass, the dynamical mass function 
behaves as
\begin{equation}
\label{Mexpl}
 M(p^2) \simeq \frac{\hat m}
{\left(\hbox{$\frac{1}{2}$} \ln\left[\frac{p^2}{\Lambda_{\rm QCD}^2}
                \right]\right)^{\gamma_m}} \,,
                \qquad p^2 \gg \Lambda_{\rm QCD}\,,
\end{equation}
where $\gamma_m = \frac{12}{33 - 2 N_f}$ is the anomalous mass dimension,
and $\hat m$ the renormalisation-point-independent current-quark
mass. Dynamical chiral symmetry breaking is manifest when, for $\hat
m=0$, one obtains $M(p^2) \neq 0$ in solving Eq.~(\ref{gendse}), which
is impossible at any finite order in perturbation theory. Analysing our
chiral limit solution we find
\begin{equation}
\label{Mchiral}
M(p^2) \simeq
\frac{2\pi^2\gamma_m}{3}\,\frac{ -\,\langle \bar q q \rangle^0 }
     {p^2\left(\case{1}{2}\ln\left[\frac{p^2}{\Lambda_{\rm QCD}^2}\right]
                \right)^{1-\gamma_m}} \,,
                \qquad p^2 \gg \Lambda_{\rm QCD}\,,
\end{equation}
with the renormalisation-point-independent chiral condensate 
$ -\,\langle \bar q q \rangle^0 = (0.227\,{\rm GeV})^3$.  
The large-$p^2$ momentum dependence of the quark mass function, both
Eq.~(\ref{Mexpl}) and Eq.~(\ref{Mchiral}), is characteristic of the QCD
renormalisation group at one-loop$^4$.

In the infrared region, we see that dynamical chiral symmetry breaking
implies more than just $ \langle\bar q q \rangle_\mu \neq 0$.  For small
momenta, $p < 1\,{\rm GeV}$, there is a qualitative and distinct
difference between light and heavy quarks.  The light-quark mass
function is very similar to the chiral limit mass function, and is
characterised by a significant infrared enhancement, a direct result of
that in the effective coupling, ${\cal G}(k^2)$. Introducing the
Euclidean constituent-quark mass, $M^E$, as the solution of
$p^2=M^2(p^2)$, the ratio: $M^E/ m(\mu)$, is a single, indicative and
quantitative measure of the nonperturbative effects of gluon-dressing on
the quark propagator. For current-quark mass values of
$m_{u/d}=3.7\,{\rm MeV}$, $m_{s}=82\,{\rm MeV}$, $m_{c}=0.59\,{\rm GeV}$
and $m_{b}=2.0\,{\rm GeV}$ at the renormalisation point 
$\mu=20\,{\rm GeV}$, we find
\begin{displaymath}
\begin{array}{l|c|c|c|c|c}
\mbox{flavour} 
        &   u/d  &   s   &  c  &  b  &  t \\ \hline
 {M^E}/{m({\mu})}
        &  150   &  10   & 2.2 &  1.4 & \to 1
\end{array}
\end{displaymath}
which clearly indicates the magnitude of this effect for light quarks.
It provides a qualitative understanding of the nature of the
``constituent-quark'' mass. The magnitude of this ratio is a signal of
the effect of the infrared enhancement of the quark-quark interaction 
on quark propagation characteristics.

For heavy quarks, the ratio $M^E/ m(\mu)$ takes a value of O$(1)$
because the current-quark mass is much larger than the mass-scale
characterising the infrared enhancement in the effective coupling,
$\Lambda_{\rm QCD}$. This means that in the spacelike region the
momentum-dependence of the heavy-quark mass function is dominated by
perturbative effects and suggests, in phenomenological applications,
that one can replace the heavy-quark mass function by a constant
mass$^5$.

Note that the both heavy and light quarks, and also the gluons, are
explicitly confined; the essence of confinement is that the propagator
of a confined particle does not have a Lehmann representation$^6$.
Therefore, quark confinement entails that there is no pole-mass, which
would be the solution of $p^2 + M^2(p^2)=0$.
\smallskip

{\bf Meson bound states}

The renormalised Bethe--Salpeter Equation [BSE] for meson bound states is
\begin{eqnarray}
\label{genbse}
\left[\Gamma_H(p;P)\right]_{tu} & = & 
        \int\frac{{\rm d}^4q}{(2\pi)^4} \, 
        [S(q_+) \Gamma_H(q;P) S(q_-)]_{sr} \,K^{rs}_{tu}(q,p;P)\,,
\end{eqnarray}
where: $H$ specifies the flavour structure of the meson; $q_+=q +
\eta_P\, P$, $q_-=q - (1-\eta_P)\, P$, with $P$ the total momentum 
of the bound state and $\eta_P$ the momentum partitioning; and
$r$,\ldots,$u$ represent colour-, Dirac- and flavour-matrix
indices. Solutions of Eq.~(\ref{genbse}) exist only for particular,
separated, timelike values of $P^2$; and the eigenvector associated with
each eigenvalue, the Bethe--Salpeter amplitude [BSA]: $\Gamma_H(p;P)$, is
the one-particle-irreducible, fully-amputated quark-meson vertex. Once
we know this BSA, we can use it to calculate meson observables such as
the decay constant and various form factors and other physical
observables. For pseudoscalar mesons, the solution of Eq.~(\ref{genbse})
has the general form$^2$
\begin{equation}
\label{genpsBSA}
\Gamma_H(q;P) = \gamma_5 \big[ i E_H(q;P) + \gamma\cdot P \, F_H(q;P) 
+ \gamma\cdot q \, G_H(q;P) + \sigma_{\mu\nu} q_\mu P_\nu \,H_H(q;P) \big] \,.
\end{equation}

In order to calculate the BSA, we must have an explicit form for the
kernel $K^{rs}_{tu}(q,p;P)$ in Eq.~(\ref{genbse}). Requiring that the
axial-vector Ward indentity\footnote{In the chiral limit, this Ward
identity relates the meson BSA to the quark propagator: the pions are
the Goldstone bosons associated with dynamical chiral symmetry
breaking.}  be preserved constrains this kernel. The ladder
approximation fulfills this constraint$^7$
\begin{equation}
\label{trunbse}
\Gamma_H(p;P) =  - \case{4}{3}\int\frac{{\rm d}^4q}{(2\pi)^4} \, 
  {\cal G}((p-q)^2)\, D_{\mu\nu}^{\rm free}(p-q)
  \gamma_\mu S(q_+)\Gamma_H(q;P) S(q_-) \gamma_\nu  \,.
\end{equation}

Numerical solutions of Eq.~(\ref{trunbse}) for increasing values of 
the current-quark mass show that the meson mass grows with the quark
mass. In Fig.~2 we present the meson mass $m_H$ as function of the
renormalisation-point-independent current-quark mass $\hat m$,
for both ($u\bar q$) ($\pi^\pm$, $K$, $D$, $B$) and (fictitious)
($q\bar q$) ($\pi^0$) pseudoscalar mesons.  
\begin{figure}[t]
\vspace*{-5mm}

\hspace*{-.5cm}
\epsfig{figure=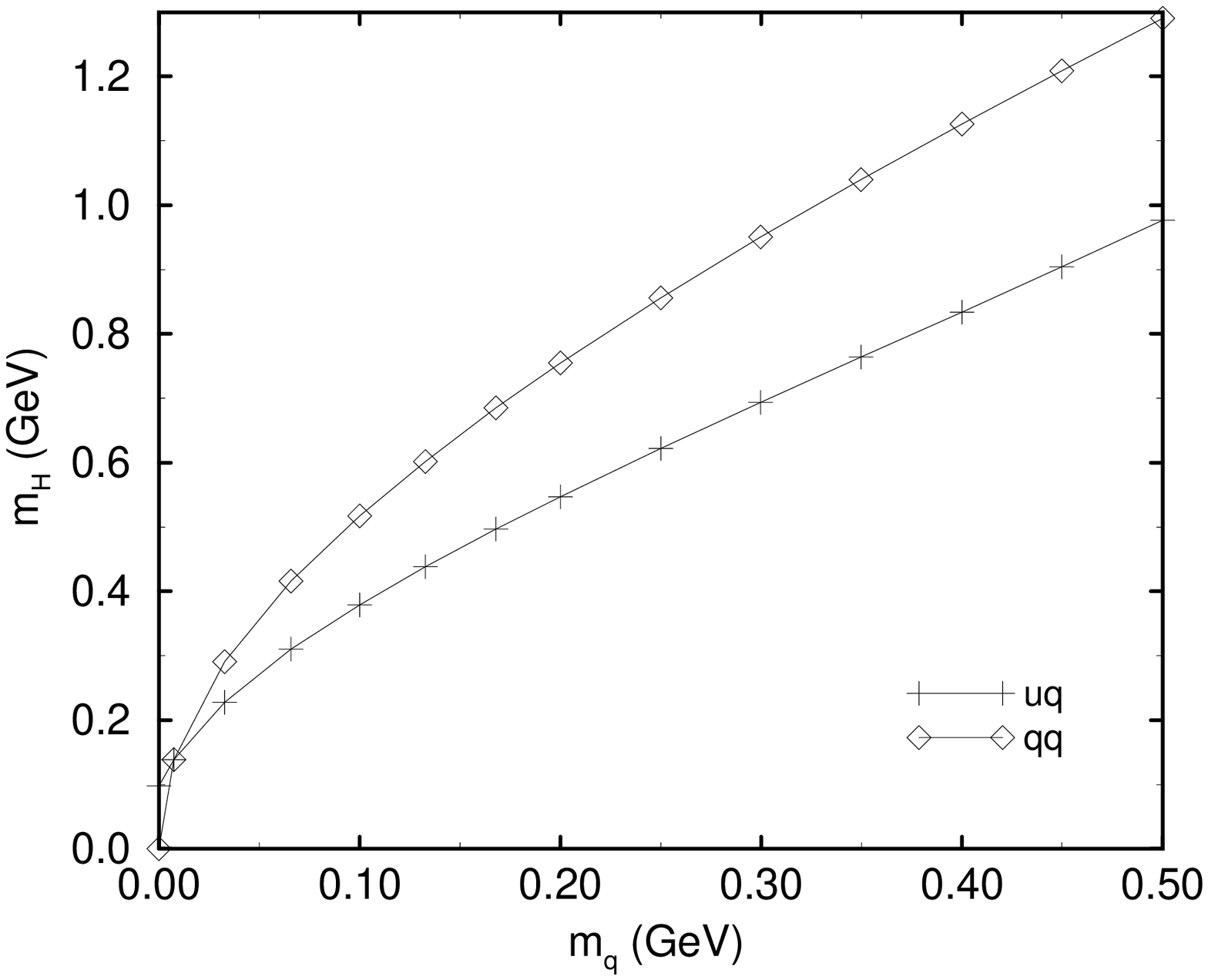,height=8cm}
\vspace*{-4cm}

\hspace*{9.5cm}
\parbox{58mm}{\small Fig.~2. The pseudoscalar meson mass for ($u\bar q$)
(plusses) and equal quark ($q\bar q$) (diamonds) bound states, as a
function of the current-quark mass $\hat m_q$.}
\vspace*{8mm}
\end{figure}
In both cases, the meson mass $m_H$ is for light quarks, proportional to
the square root of the current-quark mass, whereas for heavy quarks, the
meson mass grows linearly with the current-quark mass:
\begin{equation}
 m_H \propto \left\{ 
\begin{array}{ll}
\sqrt{\hat m_q}\,, \qquad 
        & \hat m_q < 0.15\, {\rm GeV}\, \approx\, \hat m_s \,,\\
\hat m_q       \,,       
        & \hat m_q > 0.3\, {\rm GeV}\,  \approx\, 2\,\hat m_s  \,.
\end{array}
\right.
\end{equation}

The pseudoscalar decay constant increases with quark masses, at least up
to quark masses of about $0.5\,GeV\approx 3\, \hat m_s$. The reversal of
this trend, that is dictated by the heavy quark limit $f_H \sqrt{m_H} =
{\hbox{constant}}$,$^5$ is reached at larger current-quark masses.

Although the BSA does depend on the momentum partitioning $\eta_P$ in
Eq.~(\ref{genbse}), the physical observables, such as the meson mass and
decay constant, do not, provided one takes into account the nonleading
Dirac amplitudes $F_H$ and $G_H$ in Eq.~(\ref{genpsBSA}).  These
amplitudes are also essential in order to satisfy the axial-vector Ward
identity, and are quantitatively important for physical observables$^2$.

The ultraviolet asymptotic behaviour of all amplitudes $E(p;P)$,
$F(p;P)$, $p^2\,G(p;P)$ and $p^2\,H(p;P)$, is fully determined by the
chiral-limit quark mass function: all behave like Eq.~(\ref{Mchiral}),
both for light and heavy pseudoscalar mesons. A difference between the
BSA of heavy mesons and light mesons is that heavy mesons are broader in
momentum space: they approach the asymptotic behavior at larger momenta,
and are thus narrower in coordinate space than light mesons.

\smallskip

{\bf Conclusion}

The dynamical mass function is a useful measure of nonperturbative
effects. In the infrared region, the mass functions of the light
flavours, in particular the $u$- and $d$-quarks, are very similar to the
chiral limit mass function $M_0(p^2)$. They are all characterised by a
sharp increase below $1$~GeV: the constituent quark mass is much larger
than the current-quark mass. The mass function of the heavy quarks does
not show such behaviour: nonperturbative effects are much less important
for the heavy quarks. For the $b$-quarks, and to a lesser extent the
$c$-quarks, the mass function can be approximated well by a constant:
the constituent mass is approximately the same as the current-quark
mass. One of the consequences of this difference is reflected in the
mass spectrum of the pseudoscalar mesons: the heavy meson masses
increase linearly with the current-quark masses, whereas the light meson
masses are proportional to the square root of the current-quark masses.

\smallskip

{\bf Acknowledgments} 

This work was supported by the US Department of Energy, Nuclear Physics
Division, under contract number W-31-109-ENG-38 and benefited from the
resources of the National Energy Research Scientific Computing Center.

\smallskip

\noindent
{\small
1.\ C. D. Roberts and A. G. Williams, Prog. Part. Nucl. Phys. {\bf 33} (1994)
447.\\
2.\ P. Maris and C. D. Roberts, 
nucl-th/9708029, to appear in Phys. Rev. C.\\
3.\ N. Brown and M. R. Pennington, Phys. Rev. D {\bf 39}, 2723 (1989).\\
4.\ K. Lane, Phys. Rev. {\bf 10}, 2605 (1976); H. D. Politzer,
Nucl. Phys. B {\bf 117}, 397 (1976).\\
5.\ M. A. Ivanov {\it et al.}, nucl-th/9704039, to appear in Phys. Lett. B; 
C. D. Roberts {\it et al.}, nucl-th/9710063, these proceedings.\\ 
6.\ C. D. Roberts, A. G. Williams and G. Krein, 
Int. J. Mod. Phys. A {\bf 4}, 1681 (1992); 
P. Maris, Phys. Rev. D {\bf 52}, 6087 (1995).\\
7.\ A. Bender, C. D. Roberts and L. v. Smekal, Phys. Lett. B {\bf
380}, 7 (1996).
}

\end{document}